\documentstyle[12pt,epsfig]{article}
\vspace{1 cm}
\def\lapproxeq{\lower .7ex\hbox{$\;\stackrel{\textstyle <}{\sim}\;$}}
\def\gapproxeq{\lower .7ex\hbox{$\;\stackrel{\textstyle >}{\sim}\;$}}

\begin{document}

\titlepage

\begin{flushright} RAL-TR-95-017 \\ June 1995
\end{flushright}

\begin{center}
\vspace*{2cm}
{\large{\bf Theory of Vector Meson Production}}
\end{center}

\vspace*{.75cm}
\begin{center}
J.R.\ Forshaw\footnote{Talk given in the `Diffraction and Vector Mesons'
session at the Workshop on Deep Inelastic scattering and QCD, Paris,
April 1995.}
\\
Rutherford Appleton Laboratory, \\ Chilton, Didcot OX11 0QX, England. \\
\end{center}

\vspace*{1.5cm}

\begin{abstract}
I discuss the theoretical status of the `soft' pomeron and its
place in describing generic diffractive processes. The role of
perturbative QCD (pQCD) corrections is considered, in particular in the
context of quasi-elastic vector meson production at high $Q^2$.
In those processes where short distances are dominant, the `hard' (pQCD)
pomeron is expected to reveal itself, such a process may well be that of
diffractive vector meson production at high-$t$ and I discuss this.
\end{abstract}

\newpage

\section{Introduction}
I will talk about the theoretical status of high energy diffractive/elastic
physics. As my theme I will attempt to address the questions: \lq\lq
What `tools' do we presently have?\rq\rq and \lq\lq How well do
they/should they work?\rq\rq. But, I will not attempt to discuss the
most important question: \lq\lq How are they related?\rq\rq !
I start with a review of the `soft' pomeron of Donnachie and Landshoff before
moving to the role of perturbative QCD corrections in the context of
quasi-elastic vector meson production in high-$Q^2$ $ep$ collisions at $t=0$
(when the proton usually remains intact). To conclude, I talk about a much
rarer process which ought to shed light on the perturbative (or `BFKL')
pomeron, namely that of vector meson production at high $t$ (where the
proton will usually break up).
\section{The `Soft' Pomeron}
Motivated largely by the success of the additive quark rule in describing the
ratios of the total cross sections (of light hadrons) at high energies and the
rising of the individual cross sections with increasing energy, Donnachie and
Landshoff (DL) proposed the exchange of a single Regge pole which couples
directly to on-shell valence quarks \cite{dl1}.
This simple proposition works exceedingly well for a wide range
of circumstances: total cross sections, elastic scattering at low $t$ and
quasi-elastic vector meson production at high $Q^2$ (at least at
EMC/NMC energies) are
all successfully described by a pomeron pole of trajectory $\alpha_{P}(t)
\simeq 1.08 + 0.25 t$ \cite{dl2}. We would like to understand this picture
in terms of QCD, and progress in
this direction has been made by Landshoff and Nachtmann (LN) \cite{ln}
who proposed that the pomeron is simply the exchange of two
non-perturbative gluons, see fig.1 (the blobs denote the
non-perturbative gluons).
\begin{figure}[ht]
\centerline{ \epsfig{file=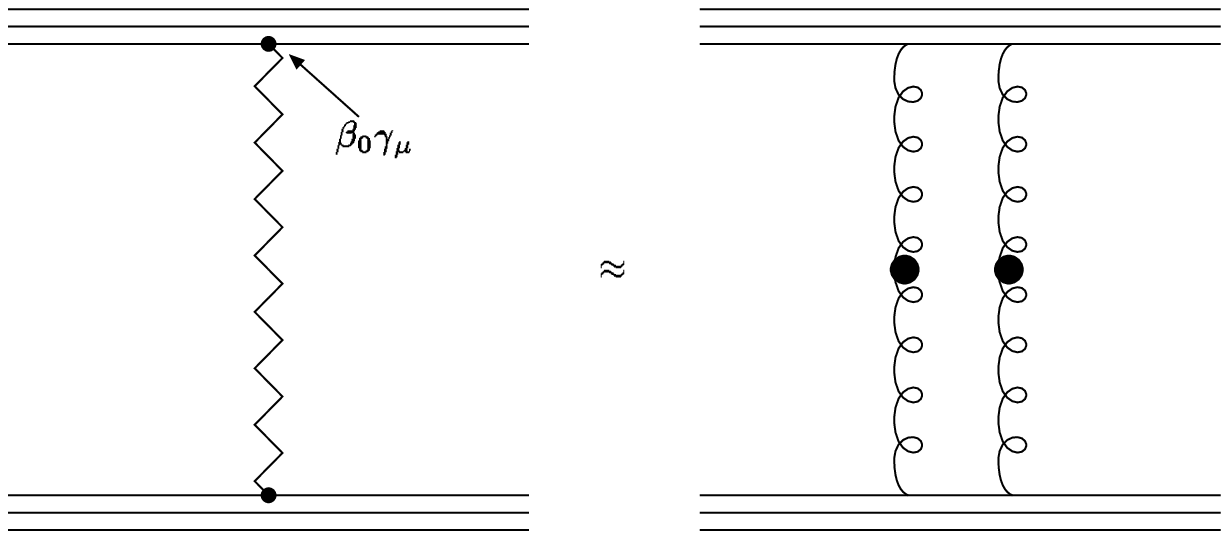,width=7.8cm}  }
\centerline{  Figure 1   }
\end{figure}
The two-point Green function that defines the
gluon propagator was shown to pick up a contribution from non-perturbative
physics which arises due to a non-zero vacuum expectation value of the gluon
condensate, $\langle G^a_{\mu \nu} G^{\mu \nu}_{a} \rangle \sim M_c^4$. The new
mass scale is related to the pomeron-quark coupling, $\beta_0$, via a
correlation length, $a$, i.e. $\beta_0 \sim M_c^4 a^5$. QCD sum
rules give $M_c$ and the DL pomeron phenomenology fixes $\beta_0$. As a
result, the correlation length is found to satisfy the inequality
$a \ll R$, where $R$ is a typical light-hadron radius.
Interpreting $a$ as the typical separation of the two non-perturbative
gluons then we can appreciate that this inequality is responsible for
guaranteeing the preservation of the additive quark rule. Unfortunately, the
LN formalism is in an Abelian theory
and rigorous contact with QCD still eludes us.

The arrival of HERA meant, for the first time, data which are not
compatible with the DL picture. The steep rise of $F_2^p(x,Q^2)$ at small $x$
\cite{hera1} and the largeness of the high-$Q^2$ quasi-elastic $\rho$
production cross section \cite{hera2} along with a similar enhancement for the
quasi-elastic photoproduction of $J/\Psi$'s \cite{hera3} are all evidence
for physics beyond the DL pomeron. In
particular they are evidence for significant perturbative corrections. Such
corrections were not unexpected, since the presence of a hard scale opens up
the phase space for perturbative corrections, e.g. $\sim \alpha_s \ln Q^2$.
A very brief word on why the diffractive contribution to the inclusive DIS
cross section appears not to rise as fast as one might naively
expect (i.e. it appears more
consistent with the `soft' pomeron approach \cite{diff}) is perhaps in order.
In fact, it is the only small $x$ deep inelastic process seen at HERA which
does not appear to contain very large perturbative corrections. The very
asymmetric partition of the longitudinal momentum of the incoming photon
between the quark and anti-quark to which it couples (in the proton rest frame)
is responsible for selecting dominantly non-perturbative configurations and
so we should not be surprised by these HERA data (see \cite{ajm,afs} for more
details).
\section{QCD corrections}
Since the perturbative calculation (of processes which involve hadrons in
the initial state) usually introduces collinear divergences it follows that
any sensible calculation must address the interface with
non-perturbative physics. Fortunately, it is known that for inclusive cross
sections these divergences can be factorised into some a priori unknown
boundary condition \cite{fac}.
\begin{figure}[ht]
\centerline{ \epsfig{file=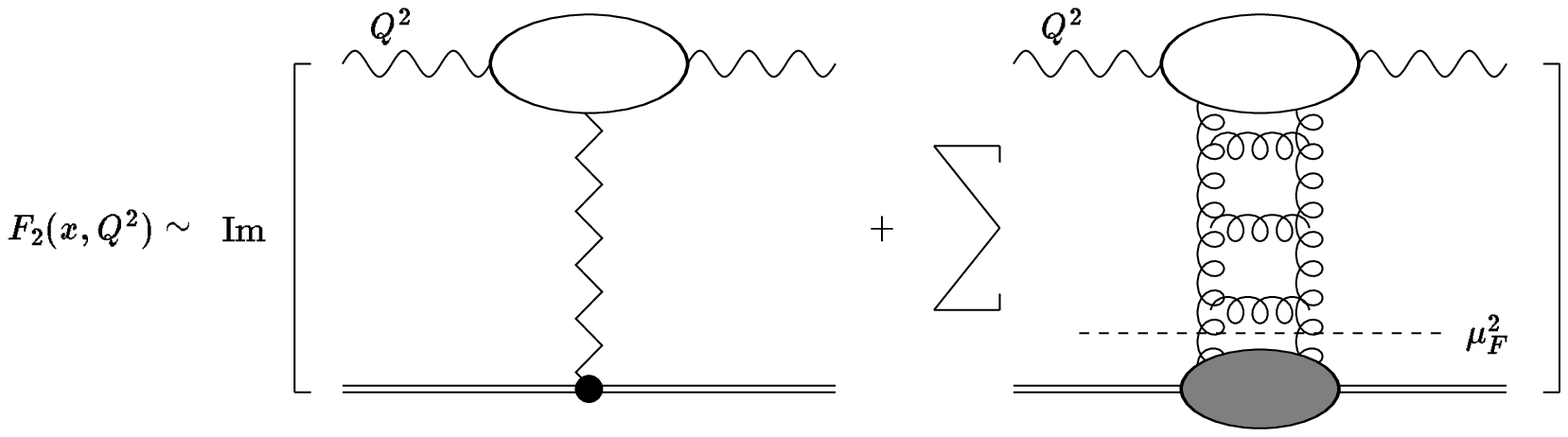,width=11.0cm}  }
\centerline{  Figure 2   }
\end{figure}
Fig.2 illustrates how the perturbative
corrections enter in a calculation of $F_2(x,Q^2)$.

What about the rapidly rising cross section for quasi-elastic vector meson
production at high $Q^2$ which has been seen in the HERA data?
In fig.3, the lowest order QCD contribution is
shown.
\begin{figure}[ht]
\centerline{ \epsfig{file=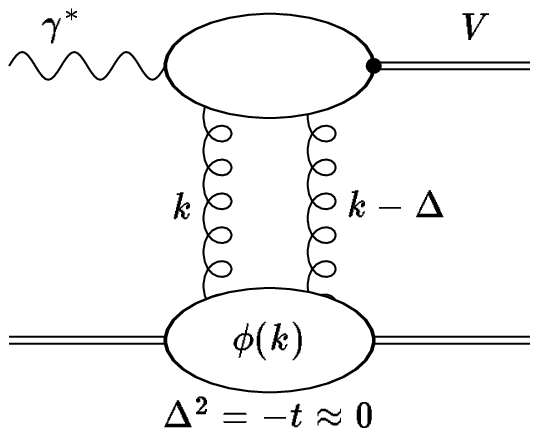,width=6.0cm}  }
\centerline{  Figure 3   }
\end{figure}
According to Ryskin \cite{ryskin},
the amplitude for scattering longitudinal
photons to produce longitudinal mesons (it is the dominant contribution) can be
written (for small enough $t$):
\begin{eqnarray}
\frac{{\mathrm{Im}}\,A(s,t)}{s} &\approx& F(t) \alpha_s \left[
\frac{\Gamma^V_{ee}m_V^3 \pi^3}{3 \alpha_{\mathrm{em}}} \right]^{1/2}
\frac{1}{Q^4} \frac{\surd Q^2}{m_V}  \nonumber \\ &\times&
\int^{\sim Q^2} \frac{dk^2}{k^4} \phi(k^2).
\end{eqnarray}
Where $F(t)$ is a form factor associated with the elastic
scattering of the proton (it is unity at $t=0$). The collinear divergence
of pQCD is present, since $\phi(k^2) \sim k^2 R^2$ at small $k^2$. It is
of the same nature as
the factorisable logarithmic divergence in $F_2$ and as such Ryskin replaces
the integral over $k^2$ with the gluon parton density, $G(x,Q^2/4)$. By
evaluating the gluon density at a scale $\sim Q^2$, the infinity of $\ln Q^2$
corrections to fig.3 are summed up. Consequently, the cross section can be
written (for $Q^2 \gg m_V^2$):
\begin{equation}
\frac{d \sigma}{dt} \sim \frac{1}{Q^6}[ G(x,Q^2/4) ]^2.
\end{equation}
Essentially the same result has been obtained by Brodksy et al \cite{brodsky}
and Nikolaev et al \cite{nikolaev}. Since the gluon density rises rapidly at
small $x$, so the cross section for $\gamma^* p \to \rho p$ rises (but twice
as fast) and we have an explanation of the HERA data. However,
we should be careful in taking this result too literally: there are huge
theoretical uncertainties in
the normalisation. These uncertainties arise since eq.(2) is derived in the
double logarithmic approximation (i.e. only $\ln Q^2 \ln 1/x$ terms are
summed up).  This approximation is necessary in order to allow us to
write the cross section simply as the square of the gluon density, evaluated
at the double leading log scales, $Q^2/W^2$ and $Q^2$.
For more discussion on the dangers associated with this expression
see Peter Landshoff's talk \cite{pvl}.
\section{The Hard `Pomeron'}
As well as a large transverse momentum phase space (which led to the large logs
in $Q^2$), at high enough CM energies there is also a large longitudinal
momentum phase space. This generates logs in $W^2/Q^2$ which lead to the much
cited BFKL corrections \cite{bfkl}. The logs exponentiate to deliver a power
law growth (in $W^2$) of total cross sections.
In fig.4, the `definitive' BFKL process is
shown: short distances are dominant and the pomeron (i.e. that object which
determines the behaviour of total cross sections at high energies) can be
described using perturbation theory.
\begin{figure}[ht]
\centerline{ \epsfig{file=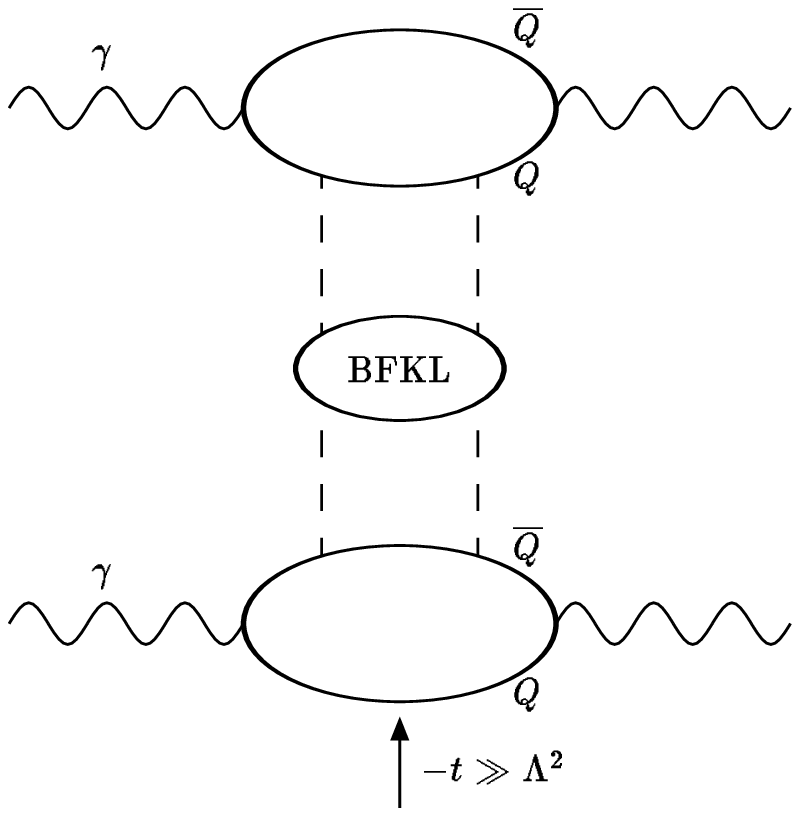,width=5.5cm}  }
\centerline{  Figure 4   }
\end{figure}
The dashed lines represent `reggeized' gluons (the bare $t$-channel gluons
having been dressed by virtual corrections). The process, $\gamma \gamma \to
\gamma \gamma$ through heavy quark loops at high-$t$, is not something
likely to be measured in the near future! Even so, one can imagine turning
down the heavy quark mass and the momentum transfer $t$
until we eventually arrive (as we must) at the DL pomeron.
Our goal must be to understand this transition. Fortunately this is
not a purely theoretical exercise,
there is a very similar process that can be measured (with decent statistics)
at HERA. This is the process, $\gamma^{(*)} p \to V + X$, where $X$ denotes
the proton dissociation \cite{fr}, see fig.5.
\begin{figure}[ht]
\centerline{ \epsfig{file=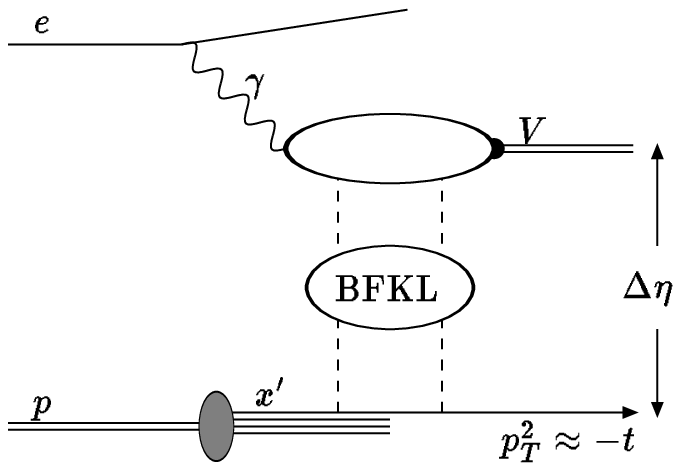,width=7.5cm}  }
\centerline{  Figure 5   }
\end{figure}
The photon need not be highly virtual,
providing $t$ is sufficiently large. Large $t$ provides a dynamical infra-red
cut-off so there is no need to worry about unknown non-perturbative physics
(the theory allows a consistency check, since one can, in principle, always
look to see how much of a contribution comes from a particular region of phase
space), it also suppresses vector dominance contributions. High $t$ is also
vital in ensuring that the simple picture of the proton dissociation shown in
fig.5 is valid, i.e. the pomeron couples to a single parton line
\cite{bartels}. The
non-perturbative physics associated with the proton bound state is then
factorised into the parton densities (which are evaluated at large $x'$
since we require a large CM energy across the pomeron and $\mu_F^2 \sim -t$ )
and are known. Another bonus is that, although a single parton is struck and
emerges to form a jet at $p_T^2 \approx -t$ we do not have to see it. The
interesting cross section, $d\sigma/dt$, can be measured by observing the decay
of the vector meson (and the scattered electron in DIS). By not requiring to
see any of the proton dissociation, we can use much more of the 820 GeV that
the proton carries into the scatter and hence pick up contributions from the
largest possible rapidity gaps that HERA can deliver (the ultimate limitation
is due to the $\sim (1-x)^5$ fall off of parton densities as $x \to 1$). This
large reach in rapidity is vital in ensuring that the whole BFKL summation is
necessary. To understand the importance of large $\Delta \eta$, recall that
the BFKL expansion is an expansion in
$$ z = \frac{3 \alpha_s}{2 \pi} \ln \left( \frac{x' W^2}{Q_H^2} \right)
\sim \alpha_s \Delta \eta,$$ where $Q_H^2$ is the hard scale
(e.g $Q_H^2 \approx -t$ for $-t \gg Q^2, m_V^2$).
Ryskin and I found that for $z \lapproxeq 0.1$, there is no need to go beyond
two-gluon exchange and that the full BFKL dynamics reveals itself only for $z
\gapproxeq 0.8$. At HERA, for $2 \le -t \le 5$ GeV$^2$, $W = 100$ GeV and $x'
\ge 0.1$ (this means that $X$ is unseen) $0.1 \le z \le 1$. So there is the
possibility to get into the most interesting regime of large $z$. The scale
invariance of the BFKL kernel means that the exchange dynamics is specified
only by $z$ and the ratio $\tau \equiv -t/(Q^2 + m_V^2)$ (providing we
assume a non-relativistic form for the vector meson wavefunction).
Going from DIS $J/\Psi$'s to photoproduction $\rho$'s corresponds to varying
$\tau$ from 0.1 to 5 which means that we can probe the
dynamics over a wide range.

The rate for this process is promising. Since, at large $z$, we feel the full
force of the BFKL power, the cross section is expected to rise rapidly with
increasing $W^2$, i.e. $\sim (W^2/Q_H^2)^{2 \omega_0}$ where $\omega_0 = 12 \ln
2 \alpha_s/\pi$. For example, for $Q^2 = 0$, $-t \ge 2$ GeV$^2$ and $W=100$ GeV
we estimate that $\sigma(\gamma p \to J/\Psi+X) \approx 5$nb and that the mean
$z \approx 0.6$. This is more than an order of magnitude larger than the
prediction based on two-gluon exchange and as such would show up rather
dramatically in the data. Note that there will be many more $\rho$'s produced
and that the high-$t$ excess should be present even in $\rho$ photoproduction.
I should note that Ryskin and I performed our calculation for the production
of transversely polarised mesons off transversely polarised photons and
assumed a small contribution from the end-points of the associated
wavefunctions. This approximation is known to be a poor one
\cite{afs,brodsky}, but our conclusions easily generalise to the case of
longitudinal photons and more realistic wavefunctions.

Not only is it interesting to study the dynamics of the exchange: understanding
the dynamics responsible for the formation of the vector meson also
challenges the theorists. The comparison of rates
for $\rho, \omega, \phi$ and $J/\Psi$ will
provide important tests. For example, DIS $\rho$ production and photoproduction
of $J/\Psi$'s can have the same $\tau$ value. Theoretically the only difference
is related to the different dynamics associated with their formation. At $t=0$,
such comparisons can be done with relative ease and puzzles such as why
$\sigma(\phi):\sigma(\rho) \approx 0.1$ (NMC \cite{nmc}) whilst theory
predicts naively 2/9 (or larger!) can be addressed.

\begin{center}
{\large\bf Aknowledgements}
\end{center}
It is a pleasure to thank the convenors of the working group and the
organizing committee for providing the opportunity to present this paper.

\end{document}